\documentclass[aps,prl,english,showpacs,superscriptaddress,twocolumn]{revtex4}
\usepackage{latexsym}
\usepackage{bm}
\usepackage{babel}
\usepackage[T1]{fontenc}
\usepackage[latin1]{inputenc}
\usepackage{mathrsfs}
\usepackage{amsmath, amssymb}
\usepackage{graphicx}
\usepackage{subfigure}

\begin{document}
\title{Estimation of a classical parameter with gaussian probes: magnetometry with
collective atomic spins}
\author{Klaus M{\o}lmer}
\affiliation{Danish National Research Foundation Center for
Quantum Optics and Department of Physics and Astronomy, University
of Aarhus, 8000 {\AA}rhus C, Denmark}
\author{Lars Bojer Madsen}
\affiliation{Department of Physics and Astronomy, University of
Aarhus, 8000 {\AA}rhus C, Denmark}

\begin{abstract}
We present a theory for the estimation of a classical magnetic
field by an atomic sample with a gaussian distribution of
collective spin components. By incorporating the magnetic field
and the probing laser field as quantum variables with gaussian
distributions on equal footing with the atoms,  we obtain a very
versatile description which is readily adapted to include probing
with squeezed light, dissipation and loss and additional
measurement capabilities on the atomic system.
\end{abstract}

\maketitle

External classical perturbations of a quantum system cause changes
in the state of the system, and a measurement of a suitable
observable provides an estimate of the strength of the
perturbation. Atoms are excellent probes for the estimation of,
e.g.,  classical electric and magnetic fields as well as for
rotations and accelerations of inertial frames. The formal
description of such ultra-sensitive measurements is quite
complicated and has only been formulated recently. The main
difficulty arises from the fact that the quantum state of the
atoms is changed due to both the interaction with the classical
perturbation and the measurement process itself which yields a
time series of stochastic outcomes. Quantum trajectory theory
\cite{Carmichael} makes it possible to simulate this stochastic
process, and descriptions are available which combine the quantum
dynamics and the parameter estimation conditioned on the detection
record \cite{Mabuchi96,Gambetta01}. Recently, the classical theory
of Kalman filters was combined with the quantum trajectory theory
\cite{stockton,geremia03}, and under the assumption that the
quantum state of the atomic system could be treated as a gaussian
state of oscillator-like degrees of freedom, and the initial
uncertainty about an applied magnetic field could also be
described by a gaussian distribution function, analytical
expressions for the precision of the estimate of the field were
derived. The analysis showed that the probing of the atomic system
squeezes the atomic observable and results in a measurement
uncertainty that decreases with time $t$ and atomic number
$N_\text{at}$ as $1/(N_\text{at}t^{3/2})$ and not as
$1/\sqrt{N_\text{at}t}$, as one might have expected from standard
counting statistics arguments.

Here, we present an alternative quantum theory for the estimation
of a $B$-field by an atomic probe. The idea is to treat both the
laser field used to probe the atoms, the atoms themselves, and the
classical $B$-field as one large quantum system. Quantum
mechanical state reduction associated with measurements then
provides directly the estimate for the expectation value and
uncertainty for the quantity of interest. Our theory arrives
easily at final estimation results, and it readily generalizes to
include decay and losses.

We will assume that a gaussian state, fully characterized by
expectation values and covariances, describes the laser field, the
atoms and the $B$-field, and we will use that the gaussian
character of the state is preserved during the evolution due to
the interactions and measurements involved. We benefit from the
considerable attention given to the transformation of gaussian
states under interactions and measurements because this class of
states permits a detailed characterization of entanglement issues
(see, e.g., \cite{Fiurasek02,GiedkeCirac,EisertPlenio,hammerer}
and references therein).

We consider a collection of atoms with a spin-1/2 ground state,
polarized along the $x$-axis. The $B$-field is assumed to point
along the $y$-axis, and it hence causes a Larmor rotation of the
atomic spins towards the $z$-axis. A linearly polarized optical
probe is transmitted through the gas. The linear probe is
decomposed into two circular components, and different couplings
to an excited state introduce a phase difference of the two field
components and cause a Faraday rotation of the polarization
proportional to the population difference between the atomic $m_z$
ground states. It is the recording of this rotation that enables
us to determine the $B$-field. The atoms are effectively described
by a collective spin operator $\bm{J} = \frac{\hbar}{2} \sum_i
\bm{\sigma}^{(i)}$, and the polarization components of the field
are described by a Stokes vector $\bm{S}$. With the initially
spin-polarized sample, and the incident field in a linearly
polarized state, we may treat $J_x$ and $S_x$ as classical
variables related to the number of atoms $N_\text{at}$ and photons
$N_\text{ph}$ via $\langle J_x \rangle =\frac{\hbar N_\text{at}}{
2}$ and $\langle S_x \rangle = \frac{\hbar N_\text{ph}}{2}$. When
the field is not too close to resonance, we may eliminate the
excited states, and the effective Hamiltonian of the atom-light
interaction can be written as  $H \propto 2 \frac {g^2}{\hbar
\Delta} J_z S_z$, with $\Delta$ the detuning from resonance. The
coupling strength between a single atom and the radiation field
(quantized within a segment of length $L=c \tau$ and area $A$) is
$g= \sqrt{\frac{\hbar \omega}{A c \tau \epsilon_0}} d/\hbar$ with
$d$ the atomic dipole moment and $\hbar \omega$ the photon energy.
It is convenient to introduce effective dimensionless position and
momentum operators for the non-classical components of the spin
and Stokes vector, $x_\text{at}= \frac{J_y}{\sqrt{ \hbar \langle
J_x \rangle}}$, $p_\text{at}=\frac{J_z}{\sqrt{\hbar \langle
 J_x \rangle}}$, $x_\text{ph} = \frac{S_y}{\sqrt{\hbar  \langle S_x \rangle}}$,
 $p_\text{ph} = \frac{S_z}{\sqrt{ \hbar \langle S_x \rangle}}$ with commutators
 $[x_i , x_j]=[p_i , p_j] = 0, [x_i, p_j] = i
 \delta_{ij}$. The perfectly polarized atomic state and the laser
field polarized along the $x$-direction correspond to the ground
state, i.e., a gaussian minimum uncertainty state of the harmonic
oscillator associated with these variable.

We assume that the probing of the atoms takes place with a
continuous wave field. Such a field can be treated as a succession
of beam segments of duration $\tau$ and with a given mean number
of photons $ N_\text{ph} = 2 \langle S_x \rangle/ \hbar = \Phi
\tau$ in each segment, with $\Phi$ the photon flux. The continuous
measurement of the field is then broken down into individual
measurements on each segment. The continuous limit is achieved
when $\tau \rightarrow 0$ and $N_\text{ph}$ in each segment gets
correspondingly small. In the limit of small $\tau$, the integral
over $\tau$ is equivalent to the application of a coarse grained
Hamiltonian given by
 $H\tau = \hbar \kappa_\tau p_\text{at} p_\text{ph}$  with dimensionless $\kappa_\tau =
\frac{2 g^2}{\Delta } \sqrt{\frac{\langle J_x \rangle}{\hbar}
\frac{\langle S_x \rangle}{\hbar}} \tau =
 \frac{2 g^2}{\Delta} \sqrt{\frac{\langle J_x \rangle}{\hbar} \frac{1}{2}
 \Phi} \tau^{3/2}$. Due to the $\tau$-dependence of $g$,
 $\kappa_\tau$ is proportional to $\sqrt{\tau}$.
 When we incorporate the $B$-field coupling to the atoms, $\beta B J_y/ \hbar $, with
 $\beta$ the atomic magnetic moment, the total effective Hamiltonian
 is given by
\begin{equation}\label{Heff}
    H \tau =\hbar \left(\kappa_\tau p_\text{at} p_{\text{ ph}} + \mu_\tau B  x_\text{at} \right),
\end{equation}
with  $\mu_\tau = \frac{1}{\hbar} \beta \sqrt{\frac{\langle J_x
\rangle }{\hbar}} \tau $.

We treat the classical $B$-field variable on equal footing with
the quantum variables. The Heisenberg equations of motion for the
column vector of the five variables ${\bm y}=
(B,x_\text{at},p_\text{at},x_\text{ph},p_\text{ph})^T$ yield ${\bm
y}(t+\tau) = \mathbf{S_\tau} {\bm y}(t)$ with the transformation
matrix
\begin{eqnarray}\label{Smatrix}
{\mathbf S_\tau} =
\left(%
\begin{array}{ccccc}
  1 & 0 & 0 & 0 & 0 \\
  0 & 1 & 0 & 0 & \kappa_\tau \\
  -\mu_\tau & 0 & 1 & 0 & 0 \\
  0 & 0 & \kappa_\tau & 1 & 0 \\
  0 & 0 & 0 & 0 & 1 \\
\end{array}%
\right).
\end{eqnarray}
The covariance matrix, defined as in
\cite{EisertPlenio,GiedkeCirac}, $\gamma_{ij} = 2 \text{Re}
\left\langle (y_i - \langle y_i \rangle ) (y_j - \langle y_j
\rangle ) \right\rangle$ then transforms as
\begin{equation}\label{gammatrans-withoutnoise}
    \bm{\gamma}(t+\tau)  = \mathbf{S_\tau} \bm{\gamma}(t)
    \mathbf{S_\tau}^T,
\end{equation}
due to the atom-light and the atom-field interaction. In the
gaussian approximation, the system is fully characterized by the
vector of expectation values $\langle {\bm y} \rangle$ and the
covariance matrix $\bm{\gamma}$. We probe the system by measuring
the Faraday rotation of the probe field, i.e., by measuring the
field observable $x_\text{ph}$. Since the photon field is an
integral part of the quantum system, this measurement will change
the state of the whole system, and in particular the covariance
matrix of the residual system of atoms and $B$-field. We denote
the covariance matrix by
\begin{eqnarray}\label{gammadecompose}
\bm{\gamma}= \left(%
\begin{array}{cc}
  \mathbf{A_\gamma} & \mathbf{C_\gamma} \\
  \mathbf{C}^T_\gamma & \mathbf{B_\gamma} \\
\end{array}%
\right),
\end{eqnarray}
where the $3 \times 3$ sub-matrix $\mathbf{A_\gamma}$ is the
covariance matrix for the variables ${\bm y }_1 = (B, x_\text{at},
p_\text{at})^T$, $\mathbf{B_\gamma}$ is the $2 \times 2$
covariance matrix for ${\bm y}_2 = (x_\text{ph}, p_\text{ph})^T$,
and $\mathbf{C_\gamma}$ is the $3 \times 2$ correlation matrix for
$\bm{y}_1$ and $\bm{y}_2^T$. An instantaneous measurement of
$x_\text{ph}$ then transforms $\mathbf{A_\gamma}$ as
\cite{Fiurasek02,GiedkeCirac,EisertPlenio}
\begin{equation}\label{Agammaupdate}
    \mathbf{A_\gamma} \mapsto  \mathbf{A}'_\gamma = \mathbf{A}_\gamma  -
    \mathbf{C_\gamma}
    (\mathbf{\pi B_\gamma} \mathbf{\pi} )^{-1} \mathbf{C}^T_\gamma,
\end{equation}
where $\mathbf{\pi} = \text{diag}(1 ,0)$, and where the inverse
denotes the Moore-Penrose pseudoinverse, as $(\mathbf{\pi B_\gamma
\pi})$ is not invertible. Equation (\ref{Agammaupdate}) is
equivalent to the result for classical gaussian random variables
derived, e.g., in \cite{Maybeck}. After the measurement, the field
part has disappeared, and a new beam segment is incident on the
atoms. This part of the beam is not yet correlated with the atoms,
and it is in the oscillator ground state, hence the covariance
matrix $\bm{\gamma}$ is updated with $\mathbf{A}'_\gamma$,
$\mathbf{C}'_\gamma$ a $3 \times 2$ matrix of zeros, and
$\mathbf{B}'_\gamma= \text{diag}(1, 1)$.

Unlike the covariance matrix update, which is independent of the
value actually measured in the optical detection, the vector
$\langle \bm{y} \rangle$ of expectation values will change in a
stochastic manner depending on the outcome of these measurements.
The outcome of the measurement on $x_\text{ph}$ after the
interaction with the atoms is random, and  the actual measurement
changes the expectation value of all other observables due to the
correlations represented by the covariance matrix. Let $\chi$
denote the difference between the measurement outcome and the
expectation value of $x_\text{ph}$, i.e., a gaussian random
variable with mean value zero and variance 1/2. The change of
$\langle \bm{y}_1\rangle$ due to the measurement is now given by:
\begin{equation}\label{expectationvalueupdate}
\langle {\bm y}_1 \rangle \mapsto  \langle {\bm y}'_1 \rangle  =
\langle {\bm y}_1 \rangle + \mathbf{C_\gamma} (\mathbf{\pi B
\pi})^{-1}(\chi,0)^T,
\end{equation}
where we use that the measurement on $x_\text{ph}$ only, leads to
the particularly simple form $(\mathbf \pi B \pi)^{-1} =
\text{diag}(B(1,1)^{-1}, 0)$, and hence the actual value of the
second entrance in the vector $(\chi,0)$ is unimportant.

The gaussian state of the system is propagated in time by repeated
use of (\ref{gammatrans-withoutnoise}) and the measurement update
formulae (\ref{Agammaupdate})-(\ref{expectationvalueupdate}). This
evolution is readily implemented numerically, and the expectation
value and our uncertainty about the value of the $B$-field are
given by the first entrance in the vector of expectation values
$\langle {\bm y}_1 \rangle  = \langle B \rangle$ and the (1,1)
entrance in the covariance matrix $\mathbf{A}_\gamma(1,1) = 2
(\Delta B)^2$.

The above discussion specifies how the parameter estimation can be
performed. In the problem at hand, the variable $x_\text{at}$ does
not couple to $B$ and $p_\text{at}$, and we are left with a closed
$2 \times 2$ system for the reduced covariance matrix  of $B$ and
$p_\text{at}$: $\mathbf{V} = [ 2 (\Delta B)^2, 2 (\Delta B
p_\text{at})^2; 2 ( \Delta p_\text{at} B)^2, 2(\Delta
p_\text{at})^2 ]$. In the limit of infinitesimally small steps the
update formulae
(\ref{gammatrans-withoutnoise})-(\ref{Agammaupdate}) translate
into a differential equation on the matrix Ricatti form
\begin{equation}\label{ricatti}
\dot{\mathbf{V}} (t) = - \mathbf{D} \mathbf{V}(t) - \mathbf{V}(t)
\mathbf{D}^T - \mathbf{V}(t) \mathbf{E} \mathbf{V}(t),
\end{equation}
with $\mathbf{D} = [ 0, 0; \mu, 0]$, $\mathbf{E} = \text{diag}(0,
\kappa^2)$, $\kappa= \kappa_\tau / \sqrt{\tau}$, and
$\mu=\mu_\tau/\tau$. We solve (\ref{ricatti}) by expressing it in
terms of two coupled linear matrix equations $\dot{\mathbf W} = -
\mathbf{D} \mathbf{W}$, $\dot{\mathbf U} = \mathbf{ E W} +
\mathbf{ D}^T \mathbf{U}$, $\mathbf{V} = \mathbf{W U}^{-1}$
\cite{stockton}, and find the analytical solution for the variance
of the magnetic field
\begin{equation}\label{Bvaranalyt}
\Delta B(t)^2 = \frac{(1+ \kappa^2 t) \Delta B_0^2 } {1+\kappa^2 t
+ \frac{2}{3} \kappa^2 \mu^2 (\Delta B_0)^2 t^3 + \frac{1}{6}
\kappa^4 \mu^2 (\Delta B_0)^2 t^4}
\end{equation}
with $\Delta B_0^2$ the initial variance. In the limit of
$\kappa^2 t \gg 1$, we have $\Delta B(t)^2 \simeq 6/(\kappa^2\mu^2
t^3)$ explicitly giving the $1/N_\text{at}$ and $1/t^{-3}$ scaling
also found in \cite{geremia03}.

The lower solid curve in  Fig.~1 shows the uncertainty of the
$B$-field as a function of time. It is worth pointing out that
compared with \cite{geremia03}, not only the spirit in which we
deal with $B$ as a quantum variable but also the formal derivation
is different. In \cite{geremia03}, the Kalman filter equation
deals with the covariance matrix for the joint estimator of the
classical $B$-field and the {\it mean value} of the atomic spin
component along the $z$-axis. The latter variance is initially
zero, because we assume that the mean value is initially known to
be zero. Our covariance matrix deals with two quantum observables,
and neither have a vanishing variance in the initial state.
\begin{figure}
\begin{center}
\includegraphics[scale=0.45]{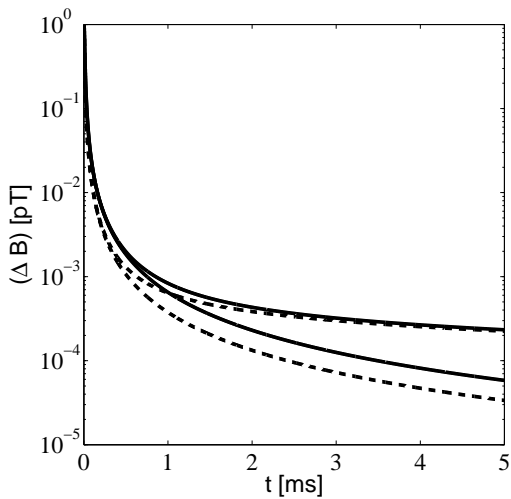}
    \end{center}
    \caption{Uncertainty of the $B$-field in pT (1\,pT = $10^{-12}$\,T)
    as function of time. We use a 2 mm$^2$ interaction
    area, $2 \times 10^{12}$ atoms, $5 \times 10^{12}$ photons s$^{-1}$, $\Delta
    B_0=1$\,pT, 1\,GHz detuning, and 852\,nm light,
    appropriate for the
    $^{133}$Cs($6S_{1/2}(F=4) - 6P_{1/2}(F=5))$ transition with
    decay rate $3.1 \times 10^7$\,s$^{-1}$ and corresponding
    atomic dipole moment $d=2.61 \times 10^{-29}$\,Cm.
    The effective couplings are $\kappa^2 =
    1.83 \times 10^{6}$\,s$^{-1}$  and $\mu=8.79 \times 10^{4}$\,(s\,
    pT)$^{-1}$. Factors of order unity related to the coupling
    matrix elements among different states of the actual
    Zeeman substructure are omitted.
    The lower curves are without inclusion of atomic decay, and
    the upper curves include atomic spontaneous emission with a
    rate $\eta =1.7577$\,s$^{-1}$. The solid (dashed) curves are
    for coherent (squeezed, $r=3$) optical probe fields (see
    text).
    }
    \label{fig:fig1}
\end{figure}

We may now go back to (\ref{expectationvalueupdate}) and derive
the stochastic differential equation
\begin{equation}
\label{stoceqB} d \langle B (t) \rangle = \sqrt{2} \kappa (\Delta
B p_\text{at})^2 dW(t)
\end{equation}
for the expectation value of the $B$-field. Here $dW(t)= \chi
\sqrt{2 dt} $ is a Wiener increment with gaussian white-noise
statistics $\langle dW(t) \rangle =0$, $\langle dW(t)^2 \rangle=
dt$. $(\Delta B p_\text{at})^2 \simeq 3/(\kappa^2 \mu t^2)$ in the
long time limit as determined by the Ricatti equation
(\ref{ricatti}), and it follows that the locking of the value of
$\langle B \rangle$,  conditioned on the measurements, takes place
predominantly in the early stages of the detection process. This
is in agreement, of course, with the rapid reduction of the
uncertainty as a function of time.

Together with the phase shift, there is a small probability that
the atoms decay by spontaneous emission from the upper probe level
to one of the two $m_z$ ground states. This occurs with a rate
    $\eta=\Phi \frac{\sigma}{A} \left( \frac{\Gamma^2/4} {\Gamma^2/4 +
\Delta^2} \right)$, where $\Gamma$ is the atomic decay width and
$\sigma=\lambda^2/( 2 \pi)$ is the resonant photon absorption
cross-section. The consequence of the decay is a loss of spin
polarization. If every atom has a probability $\eta_\tau = \eta
\tau$ to decay in time $\tau$ with equal probability into the two
ground states, the collective mean spin vector is reduced by the
corresponding factor $\langle \bm{J}\rangle \rightarrow \langle
\bm{J}\rangle (1-\eta_\tau)$. When the classical $x$-component is
reduced this leads to a reduction with time of the coupling
strengths $\kappa_\tau \mapsto \kappa_\tau \sqrt{1- \eta_\tau} $
and $\mu_\tau \mapsto \mu_\tau \sqrt{1-\eta_\tau} $, which was
also discussed in \cite{geremia03,hammerer}, and the vector of
expectation values evolves as $\langle \bm{y}(t+\tau) \rangle  =
\mathbf{L_\tau S_\tau } \langle \bm{y}(t) \rangle$ with
$\mathbf{L_\tau} =
\text{diag}(1,\sqrt{1-\eta_\tau},\sqrt{1-\eta_\tau}, 1, 1)$.

The fraction $\eta_\tau$ of  atoms that have decayed represents a
loss of collective squeezing because its correlation with the
other atoms is lost, whereas it still provides a contribution
$\hbar^2/4$ per atom to the collective spin variance. The mean
value of $J_z^2$ can be expressed in terms of the mean values of
the $N_\text{at}(N_\text{at}-1)$ atomic correlations $\sigma_z^i
\sigma_z^j$, and counting terms, we find that $\langle
J_z^2\rangle \rightarrow (1-\eta_\tau)^2 \langle J_z^2\rangle +
(\hbar^2 N_\text{at}/4)(1-(1-\eta_\tau)^2)$. Translating this and
similar expressions for $J_y^2$ and $J_xJ_y$ into the appropriate
formulae for the effective position and momentum observables,
(\ref{gammatrans-withoutnoise}) generalizes to
\begin{equation}\label{gammatrans-withnoise}
\bm{\gamma}(t+\tau)= \mathbf{L_\tau} \mathbf{S_\tau}
\bm{\gamma}(t) \mathbf{S}^T_\tau \mathbf{L_\tau} + \frac{\hbar
N_\text{at}}{\langle J_x(t)\rangle}  \mathbf{M_\tau},
\end{equation}
for $\eta_\tau \ll 1$ with $\mathbf {M_\tau}=
\text{diag}(0,\eta_\tau,\eta_\tau,0,0)$. The prefactor $\hbar
N_\text{at}/\langle J_x(t)\rangle$ initially attains the value 2,
and increases by the factor $(1-\eta_\tau)^{-1}$ in each time step
$\tau$. The effects of measurements on the covariance matrix and
the expectation value vector are obtained as in the case without
noise, and  for $\eta_\tau = 0$ we regain the noise-less case.

The upper solid curve in Fig.~1 shows the results of the
measurement when noise is taken into account. The covariance
matrix makes the atomic probe broader, and simultaneously, the
effective coupling of the atoms to the light field and to the
$B$-field is reduced, so that the knowledge acquired in the
initial detection stages is preserved but the uncertainty $\Delta
B$ does not decrease indefinitely.

The value of $B$ is estimated by the polarization rotation of the
optical field, and it is natural to enquire whether the use of
polarization squeezed light with a smaller variance of
$x_\text{ph}$ may be utilized to improve the estimate. To analyze
this proposal, we go back to our update formulae and represent
each new segment of the incident field with gaussian variances
$\mathbf{B}'_\gamma=\text{diag}(1/r,r)$, and leave all other
operations unchanged. The result is a reduction of the variance of
our estimate, shown as the dashed curves in Fig.~1. The upper
dashed curve is for the case when noise is included. The Ricatti
equation can be solved in the noise-less case, and the only change
of the result in (\ref{Bvaranalyt}) is that all occurrences of
$\kappa^2$ are replaced by $r\kappa^2$. In the long time limit,
the estimate is improved by the factor $1/r$. Since the optical
field is not squeezed if the time segments $\tau$ are shorter than
the squeezing bandwidth $\Omega$, we rely on a separation of time
scales $\Omega^{-1} \ll \tau \ll \mu^{-1}, \kappa^{-2}$ for the
above update formulae to be valid, and for the Ricatti equation to
provide a precise analytical solution. For the parameters used in
Fig.~1, the squeezing bandwidth should be larger than 10\,MHz.
Effects of finite squeezing bandwidth will be analyzed elsewhere.

We can improve our estimate by noting that the covariance matrix
describes correlations between the atomic observables and the
$B$-field, and the uncertainty in the measurement is linked with
the uncertainty of the atomic observable $x_\text{at}$. After the
optical probing it is in principle possible to perform a
destructive (Stern-Gerlach) measurement of this atomic variable.
This can of course only be done once. The formal treatment of
measurements in (\ref{Agammaupdate}) also applies when the atomic
component is being measured, and we can readily determine the new
variance on the $B$-field estimate. From the Ricatti equations we
know the covariance matrix $\mathbf{A}_{\gamma}$ analytically, and
assuming an atomic measurement at time $t$, we obtain $\Delta
B_\text{SG}(t)^2 = \frac{(\Delta B_0)^2} {1 + 2 \mu^2 (\Delta
B_0)^2 t^2 + \frac{2}{3} \kappa^2 \mu^2 (\Delta B_0)^2 t^3}$. This
variance is smaller than $(\Delta B(t))^2$ from
(\ref{Bvaranalyt}), and in the long-time limit the variance is
reduced by a factor of 4.

In summary, we have described a theory for the estimation of a
classical $B$-field by an atomic ensemble with a gaussian
distribution of collective spin components. Our theory makes use
of results obtained in the study of classification and
characterization of entanglement in continuous-variable systems
\cite{EisertPlenio}. In general, the gaussian ansatz holds for
Hamiltonians which are at most second order polynomials  in the
canonical variables, and the gaussian character of a system is
maintained under physical operations which are implemented using
linear optical elements and homodyne measurements
\cite{GiedkeCirac}. It is clearly convenient to have a unified
formalism that deals with both the probing field, the atomic
probe, and the unknown $B$-field, and which bypasses the need for
separate probabilistic arguments to yield the final estimator. The
treatment of the unknown $B$-field as a quantum variable is not
incompatible with our assumption that it is a classical parameter.
We may imagine a canonically conjugate variable to $B$ having an
uncertainty much larger than required by Heisenberg's uncertainty
relation and/or additional physical systems, entangled with the
$B$-variable, in which cases the $B$-distribution is indeed
incoherent and "classical". Also, one may argue that all classical
variables are actually quantum mechanical variables for which a
classical description suffices, and hence our theory provides the
correct estimator: quantum mechanics dictates that the quantum
state provides all the available knowledge about a system, and any
estimator providing a tighter bound hence represents additional
knowledge equivalent to a hidden variable, and this is excluded by
quantum theory. It is of course crucial that our measurement
scheme corresponds to a quantum non-demolition (QND) measurement,
i.e., we assume that there is not a free evolution of the
$B$-field induced by its conjugate variable which may thus remain
unspecified. It is also this QND property of the measurement
scheme that implies the monotonic reduction of $\Delta B$ which is
consistent with the classical parameter estimation (we can not
unlearn what we have already learnt about $B$), unlike, e.g., the
uncertainty of the atomic $x_\text{at}$ variable which must
increase when $\Delta p_\text{at}$ is reduced and when the atoms
undergo spontaneous decay.

We expect extensions of the present theory to be applicable to the
description of a variety of experiments aiming at ultra-high
precision, including, e.g., atomic clocks, studies of parity
violation, and the detection  of gravitational waves.

L.B.M. is supported by the Danish Natural Science Research Council
(Grant No. 21-03-0163).


\begin{thebibliography}{10}
\expandafter\ifx\csname
natexlab\endcsname\relax\def\natexlab#1{#1}\fi
\expandafter\ifx\csname bibnamefont\endcsname\relax
  \def\bibnamefont#1{#1}\fi
\expandafter\ifx\csname bibfnamefont\endcsname\relax
  \def\bibfnamefont#1{#1}\fi
\expandafter\ifx\csname citenamefont\endcsname\relax
  \def\citenamefont#1{#1}\fi
\expandafter\ifx\csname url\endcsname\relax
  \def\url#1{\texttt{#1}}\fi
\expandafter\ifx\csname
urlprefix\endcsname\relax\def\urlprefix{URL }\fi
\providecommand{\bibinfo}[2]{#2}
\providecommand{\eprint}[2][]{\url{#2}}

\bibitem[{\citenamefont{Carmichael}(1993)}]{Carmichael}
\bibinfo{author}{\bibfnamefont{H.}~\bibnamefont{Carmichael}},
  \emph{\bibinfo{title}{An Open Systems Aproach to Quantum Optics}}
  (\bibinfo{publisher}{(Springer-Verlag, Berlin)}, \bibinfo{year}{1993}).

\bibitem[{\citenamefont{Mabuchi}(1996)}]{Mabuchi96}
\bibinfo{author}{\bibfnamefont{H.}~\bibnamefont{Mabuchi}},
  \bibinfo{journal}{Quant. Semiclass. Opt.} \textbf{\bibinfo{volume}{8}},
  \bibinfo{pages}{1103} (\bibinfo{year}{1996}).

\bibitem[{\citenamefont{Gambetta and Wiseman}(2001)}]{Gambetta01}
\bibinfo{author}{\bibfnamefont{J.}~\bibnamefont{Gambetta}} \bibnamefont{and}
  \bibinfo{author}{\bibfnamefont{H.~M.} \bibnamefont{Wiseman}},
  \bibinfo{journal}{Phys. Rev. A} \textbf{\bibinfo{volume}{64}},
  \bibinfo{pages}{042105} (\bibinfo{year}{2001}).

\bibitem[{\citenamefont{Stockton et~al.}(2003)\citenamefont{Stockton, Geremia,
  Doherty, and Mabuchi}}]{stockton}
\bibinfo{author}{\bibfnamefont{J.~K.} \bibnamefont{Stockton}},
  \bibinfo{author}{\bibfnamefont{J.~M.} \bibnamefont{Geremia}},
  \bibinfo{author}{\bibfnamefont{A.~C.} \bibnamefont{Doherty}},
  \bibnamefont{and} \bibinfo{author}{\bibfnamefont{H.}~\bibnamefont{Mabuchi}},
  \bibinfo{journal}{quant-ph/0309101}  (\bibinfo{year}{2003}).

\bibitem[{\citenamefont{Geremia et~al.}(2003)\citenamefont{Geremia, Stockton,
  Doherty, and Mabuchi}}]{geremia03}
\bibinfo{author}{\bibfnamefont{J.}~\bibnamefont{Geremia}},
  \bibinfo{author}{\bibfnamefont{J.~K.} \bibnamefont{Stockton}},
  \bibinfo{author}{\bibfnamefont{A.~C.} \bibnamefont{Doherty}},
  \bibnamefont{and} \bibinfo{author}{\bibfnamefont{H.}~\bibnamefont{Mabuchi}},
  \bibinfo{journal}{Phys. Rev. Lett.} \textbf{\bibinfo{volume}{91}},
  \bibinfo{pages}{250801} (\bibinfo{year}{2003}).

\bibitem[{\citenamefont{Fiur\'{a}\v{s}ek}(2002)}]{Fiurasek02}
\bibinfo{author}{\bibfnamefont{J.}~\bibnamefont{Fiur\'{a}\v{s}ek}},
  \bibinfo{journal}{Phys. Rev. Lett.} \textbf{\bibinfo{volume}{89}},
  \bibinfo{pages}{137904} (\bibinfo{year}{2002}).

\bibitem[{\citenamefont{Giedke and Cirac}(2002)}]{GiedkeCirac}
\bibinfo{author}{\bibfnamefont{G.}~\bibnamefont{Giedke}} \bibnamefont{and}
  \bibinfo{author}{\bibfnamefont{J.~I.} \bibnamefont{Cirac}},
  \bibinfo{journal}{Phys. Rev. A} \textbf{\bibinfo{volume}{66}},
  \bibinfo{pages}{032316} (\bibinfo{year}{2002}).

\bibitem[{\citenamefont{Eisert and Plenio}(2003)}]{EisertPlenio}
\bibinfo{author}{\bibfnamefont{J.}~\bibnamefont{Eisert}} \bibnamefont{and}
  \bibinfo{author}{\bibfnamefont{M.~B.} \bibnamefont{Plenio}},
  \bibinfo{journal}{quant-ph/0312071}  (\bibinfo{year}{2003}).

\bibitem[{\citenamefont{Hammerer et~al.}(2003)\citenamefont{Hammerer,
  M{\o}lmer, Polzik, and Cirac}}]{hammerer}
\bibinfo{author}{\bibfnamefont{K.}~\bibnamefont{Hammerer}},
  \bibinfo{author}{\bibfnamefont{K.}~\bibnamefont{M{\o}lmer}},
  \bibinfo{author}{\bibfnamefont{E.~S.} \bibnamefont{Polzik}},
  \bibnamefont{and} \bibinfo{author}{\bibfnamefont{J.~I.} \bibnamefont{Cirac}},
  \bibinfo{journal}{quant-ph/0312156}  (\bibinfo{year}{2003}).

\bibitem[{\citenamefont{Maybeck}(1979)}]{Maybeck}
\bibinfo{author}{\bibfnamefont{P.~S.} \bibnamefont{Maybeck}},
  \emph{\bibinfo{title}{Stochastic Models, Estimation and Control. Volume 1}}
  (\bibinfo{publisher}{Academic Press: New York}, \bibinfo{year}{1979}).

\end{thebibliography}

\end{document}